\documentstyle[preprint,aps,epsfig]{revtex}

\tightenlines

\begin{document}

\title{Strange quark matter in a chiral SU(3) quark 
mean field model}
\author{P. \  Wang, V. \ E. \ Lyubovitskij,  
Th. \ Gutsche, and Amand \, Faessler}
\address{Institut f\"ur Theoretische Physik, Universit\"at
T\"ubingen, Auf der Morgenstelle 14,  \\
D-72076 T\"ubingen, Germany}
 
\maketitle
 
\vskip.5cm
 
\begin{abstract}
We apply the chiral SU(3) quark mean field model to investigate 
strange quark matter. The stability of strange quark matter with 
different strangeness fraction is studied. 
The interaction between quarks and vector mesons destabilizes the 
strange quark matter. If the strength of the vector coupling is the 
same as in hadronic matter, strangelets can not be formed. 
For the case of $\beta$ equilibrium, there is no strange 
quark matter which can be stable against hadron emission even 
without vector meson interactions.
\end{abstract}
 
\vskip.5cm


\vskip.5cm
 

\section{Introduction}

Strange quark matter has attracted a lot of interest since Witten 
suggested that it could be absolutely stable even at zero temperature 
and pressure \cite{Witten}. The investigation of such a possibility is 
relevant not only for high energy physics, but also for astrophysics. 
For example, the core of a neutron star may be composed of quark 
matter. The possible existence of strange stars which are made entirely 
of deconfined u, d and s quarks is one of the most intriguing aspects 
of modern astrophysics. There have been some reports of events with A 
$\simeq$ 350-500 and Z $\simeq$ 10-20 in cosmic ray experiments 
\cite{Kasuya}-\cite{Capdevielle}, the so-called exotic cosmic ray 
events. Also, recent studies have shown that X-ray burst sources are 
likely strange star candidates \cite{Bombaci}-\cite{Li}. It is also 
interesting to produce strange quark matter (strangelets) in the 
laboratory because they could serve as a signature of the formation of 
the quark-gluon-plasma which is a direct demonstration of 
QCD \cite{Greiner1}-\cite{Greiner3}. Many ultrarelativistic heavy-ion 
collision experiments at Brookhaven and CERN \cite{Barrette} are 
proposed to search for (meta)stable lumps of such kind of strangelets. 
Recently, Ardouin et al.\cite{Ardouin} presented a novel method which 
can be applied to characterize the possible existence of a strange 
quark matter distillation process in heavy-ion collisions. Up to now, 
there is no experiment which confirms the existence of strangelets. 
For example, the E864 collaboration found that there is no evidence 
for strangelet production in 11.5GeV/c per nucleon Au+Pb collisions 
\cite{Armstrong}. 

Besides the experimental efforts, there are also a lot of theoretical investigations 
of the stability of strange quark matter. The earliest discussions are 
based on the MIT bag model \cite{Chodos} which assumes that quarks are 
confined by a phenomenological bag. Within the bag quarks are 
asymptotically free. Calculations \cite{Farhi} within this model 
indicate that there is a range of parameters in which strange quark 
matter is absolutely stable, i.e. the energy per baryon is less than 
$930$ MeV. The stability of strange quark matter with finite volume 
(strangelets) was also discussed in the MIT bag model. 
Berger and Jaffe \cite{Berger} discussed the surface correction for the 
strangelets, where they found that the surface tension destabilizes 
strangelets. The curvature contribution was considered by 
Madsen \cite{Madsen} which is dominant for strangelets with small 
baryon numbers. Though the bag model is simple, it is an incomplete 
description of confinement. Results from lattice calculations \cite{Ukawa} 
show that quark matter does not become asymptotically free 
and some hadronic degrees of freedom remain within the quark matter 
immediately after the phase transition. Fowler, Raha and Weiner 
\cite{Fowler} suggested another description of the confinement 
mechanism via the introduction of a density-dependent quark mass. 
This quark mass-density-dependent (QMDD) model was first employed to 
study the properties of ordinary quark matter \cite{Fowler} and then 
applied to the investigation of strange quark matter 
\cite{Chakrabarty}-\cite{Peng}. As was pointed in our recent 
paper \cite{Wang1}, their thermodynamic treatment was not correct. We 
reconsidered strange quark matter in the self-consistent quark mass 
density dependent model and found a region of parameters in which the 
strange quark matter is absolutely stable. 

In the QMDD model, the concept of a density dependent quark mass has 
no dynamical origin. In recent years, some approaches for strange quark 
matter based on dynamical models were developed. Alberico et al. 
\cite{Alberico} utilized the color dielectric model to calculate the 
energy per baryon of strange quark matter. They found that while the 
double minimum version of the color dielectric model allowed the 
existence of strangelets, the single minimum version of this model 
excluded the possibility. Stability of strange quark matter was also 
investigated using the effective 4-quark interactions \cite{Jaminon}, 
the SU(3) Nambu-Jona-Lasinio (NJL) model with and without 4-quark 
vector type interactions \cite{Mishustin,Buballa}. In studying hadronic 
matter, we proposed a chiral SU(3) quark mean field model. This chiral 
quark model was applied to investigate the properties of strange 
hadronic matter and multi-strange hadronic systems \cite{Wang2,Wang3}. 
In this paper, we want to use this model to discuss the stability of 
strange quark matter. The difference between quark and hadronic matter 
is that in quark matter, the u, d, s quarks are deconfined and not 
combined into baryons by the confining potential. 

The paper is organized as follows. In section II, we introduce the 
basic model features. We apply the model to investigate strange quark 
matter in section III. The numerical calculations are discussed in 
section IV. Finally, main conclusions are drawn in section V. 

\section{The model}

Our considerations are based on the chiral SU(3) quark mean field model 
(for details see Refs. \cite{Wang2,Wang3}). For completeness, we 
introduce the main concepts of the model in this section. In the chiral 
limit, the quark field $q$ can be split into left and right-handed parts $q_L$ 
and $q_R$: $q \, = \, q_L \, + \, q_R$. Under $SU(3)_L\times SU(3)_R$ 
they transform as 
\begin{equation}
q_L' \, = \, L \, q_L, ~~~~~ q_R' \, = \, R \, q_R \, . 
\end{equation}
The spin-0 mesons are written in the compact form 
\begin{equation}
M(M^+) = \Sigma\pm i\Pi = \frac1{\sqrt{2}} \sum_{a=0}^8 
\left(\sigma^a\pm i\pi^a\right) \lambda^a, 
\end{equation}
where $\sigma^a$ and $\pi^a$ are the nonets of scalar and pseudoscalar 
mesons, respectively, $\lambda^a (a=1,...,8)$ are the Gell-Mann 
matrices, and $\lambda^0=\sqrt{\frac{2}{3}} \, I$. 
Plus and minus sign correspond to $M$ and $M^+$. 
Under chiral SU(3) transformations, $M$ and $M^+$ transform as 
$M \to M' = LMR^+$ and $M^+ \to M^{+'} = RM^+L^+$. As for the spin-0 
mesons, the spin-1 mesons are set up in a similar way as 
\begin{equation} 
l_\mu(r_\mu)=\frac12\left(V_\mu\pm A_\mu\right)=\frac1{2\sqrt{2}} 
\sum_{a=0}^8\left(v_\mu^a\pm a_\mu^a\right)\lambda^a. 
\end{equation}
They transform as $l_\mu \to l_\mu' = Ll_\mu L^+$, $r_\mu'=Rr_\mu R^+$. 
These matrices can be written in a form where the physical states are 
explicit. For the scalar and vector nonets, the expressions are 
\begin{eqnarray}
\Sigma=\frac1{\sqrt{2}}\sum_{a=0}^8\sigma^a\lambda^a=\left( 
\begin{array}{lcr}
\frac1{\sqrt{2}}\left(\sigma+a_0^0\right) & a_0^+ & K^{*+} \\
a_0^- & \frac1{\sqrt{2}}\left(\sigma-a_0^0\right) & K^{*0} \\
K^{*-} & \bar{K}^{*0} & \zeta
\end{array} \right),
\end{eqnarray}
\begin{eqnarray}
V_\mu=\frac1{\sqrt{2}}\sum_{a=0}^8
v_\mu^a\lambda^a=\left(
\begin{array}{lcr}
\frac1{\sqrt{2}}\left(\omega_\mu+\rho_\mu^0\right) & 
\rho_\mu^+ & K_\mu^{*+} \\
\rho_\mu^- & \frac1{\sqrt{2}}\left(\omega_\mu-\rho_\mu^0\right) 
& K_\mu^{*0} \\ K_\mu^{*-} & \bar{K}_\mu^{*0} & \phi_\mu 
\end{array} \right).
\end{eqnarray}
Pseudoscalar and pseudovector nonet mesons can be written in the 
same way. 

The total effective Lagrangian for the description of 
strange quark matter is given by:
\begin{eqnarray} 
{\cal L}_{\rm eff} \, = \, {\cal L}_{q0} \, + \, {\cal L}_{qM} 
\, + \, {\cal L}_{\Sigma\Sigma} \, + \, {\cal L}_{VV} \, + \, 
{\cal L}_{\chi SB} \, + \, {\cal L}_{\Delta m_s} \, + \, {\cal L}_{h}.
\end{eqnarray}
It contains the free part for massless quarks 
${\cal L}_{q0} = \bar q \, i\gamma^\mu \partial_\mu \, q$, 
the quark-meson field interaction term
\begin{eqnarray}
{\cal L}_{qM} &=& g_s\left( \bar q_L M q_R+\bar q_R M^+ q_L \right) 
- g_v \left( \bar q_L \gamma^\mu l_\mu q_L + \bar q_R \gamma^\mu 
r_\mu q_R \right) , 
\end{eqnarray}
the chiral-invariant scalar meson ${\cal L}_{\Sigma\Sigma}$ and vector 
meson ${\cal L}_{VV}$ self-interaction terms in the mean field 
approximation \cite{Wang2,Papazoglou} 
\begin{eqnarray} 
{\cal L}_{\Sigma\Sigma} &=& -\frac{1}{2} \, k_0\chi^2 
\left(\sigma^2+\zeta^2\right)+k_1
\left(\sigma^2+\zeta^2\right)^2+k_2\left(\frac{\sigma^4}2
+\zeta^4\right)+k_3\chi\sigma^2\zeta \label{scalar}\nonumber \\
&&-k_4\chi^4-\frac14\chi^4ln\frac{\chi^4}{\chi_0^4}
+\frac{\delta}3\chi^4ln\frac{\sigma^2\zeta}{\sigma_0^2\zeta_0},\\
{\cal L}_{VV}&=&\frac{1}{2} \, \frac{\chi^2}{\chi_0^2} \left(
m_\omega^2\omega^2+m_\rho^2\rho^2+m_\phi^2\phi^2\right)+g_4 
\left(\omega^4+6\omega^2\rho^2+\rho^4+2\phi^4\right) \label{vector} , 
\end{eqnarray}
where $\delta = 6/33$; $\sigma_0$, $\zeta_0$ and $\chi_0$ are the 
vacuum values of the mean fields $\sigma$,  $\zeta$ and $\chi$ 
and the three terms 
${\cal L}_{\chi SB}$, ${\cal L}_{\Delta m_s}$ and ${\cal L}_h$ which 
explicitly break the chiral symmetry.  
Chiral symmetry requires the following basic 
relations for the quark-meson coupling constants: 
\begin{eqnarray}
\frac{g_s}{\sqrt{2}} 
&=& g_{a_0}^u = -g_{a_0}^d = g_\sigma^u = g_\sigma^d = \ldots = 
\frac{1}{\sqrt{2}}g_\zeta^s,
~~~~~g_{a_0}^s = g_\sigma^s = g_\zeta^u = g_\zeta^d = 0 \, ,\\
\frac{g_v}{2\sqrt{2}} 
&=& g_{\rho^0}^u = -g_{\rho^0}^d = g_\omega^u = g_\omega^d = \ldots = 
\frac{1}{\sqrt{2}}g_\phi^s,
~~~~~g_\omega^s = g_{\rho^0}^s = g_\phi^u = g_\phi^d = 0 .
\end{eqnarray}
Note, the values of $\sigma_0$, $\zeta_0$ and $\chi_0$ are determined 
later from Eqs. (\ref{eq_gsigma})-(\ref{eq_chi}). Particularly, 
the parameters $\sigma_0$ and $\zeta_0$ are expressed through 
the pion ($F_\pi$ = 93 MeV) and the kaon ($F_K$ = 115 MeV) leptonic 
decay constants as: 
\begin{eqnarray}
\sigma_0 = - F_\pi    \hspace*{1cm} \hspace*{1cm} 
\zeta_0  = \frac{1}{\sqrt{2}} ( F_\pi - 2 F_K) 
\end{eqnarray}
The Lagrangian ${\cal L}_{\chi SB}$ generates the nonvanishing masses 
of pseudoscalar mesons 
\begin{equation} 
{\cal L}_{\chi SB}=\frac{\chi^2}{\chi_0^2}\left[m_\pi^2f_\pi\sigma + 
\left(\sqrt{2}m_K^2f_K-\frac{\sqrt{2}}m_\pi^2f_\pi\right)\zeta\right],
\end{equation} 
leading to a nonvanishing divergence of the axial 
currents which satisfy the PCAC relations for $\pi$ and $K$ mesons. 
Scalar mesons obtain the masses by spontaneous breaking of 
the chiral symmetry in the Lagrangian (\ref{scalar}). The masses of 
$u$, $d$ and $s$ quarks are generated by the vacuum expectation values 
of the two scalar mesons $\sigma$ and $\zeta$. To obtain the 
correct constituent mass of the strange quark, an additional mass 
term should be added: 
\begin{eqnarray} 
{\cal L}_{\Delta m_s} = - \Delta m_s \bar q S q
\end{eqnarray}
where $S \, = \, \frac{1}{3} \, \left(I - \lambda_8\sqrt{3}\right) = 
{\rm diag}(0,0,1)$ is the strangeness quark matrix.  
Finally, the  quark masses are given by 
\begin{eqnarray}
m_u=m_d=-\frac{g_s}{\sqrt{2}}\sigma_0, ~~~ 
m_s=-g_s \zeta_0 + \Delta m_s, 
\end{eqnarray} 
The parameters $g_s = 4.76$ and $\Delta m_s = 29$ MeV are determined from $m_q=313$ MeV 
and $m_s=490$ MeV. 
In order to obtain reasonable hyperon potentials in hadronic matter 
we include the additional coupling between strange quarks and scalar 
mesons $\sigma$ and $\zeta$ \cite{Wang2}. This term is expressed as
\begin{eqnarray}
{\cal L}_h \, = \, (h_1 \, \sigma \,  + \, h_2 \, \zeta) \, \bar{s} s \, .  
\end{eqnarray}

\section{Application to strange quark matter}

Now we apply the model to investigate strange quark matter. We begin 
with the thermodynamical potential because all other quantities such 
as energy per volume and pressure can be obtained from it. 
The thermodynamical potential is defined as 
\begin{eqnarray}
\hspace*{-.4cm} 
\Omega=\sum_{\tau=q,e}\frac{-2k_BT\gamma_\tau}{(2\pi)^3}\int_0^\infty 
d^3k\left\{ {\rm ln} 
\left(1+e^{-(E_\tau^*(k)-\nu_\tau)/k_BT}\right)+
{\rm ln} \left(1+e^{-(E_\tau^*(k)+\nu_\tau)/k_BT}\right)\right\} - 
{\cal L}_M,
\end{eqnarray}
where $E_\tau^*(k)=\sqrt{m_\tau^{*2}+k^2}$,  
$\gamma_\tau$ is 3 for quarks and 1 for electrons and 
${\cal L}_M$ is the meson interaction including the scalar meson 
self-interaction ${\cal L}_{\Sigma\Sigma}$, the vector meson 
self-interaction ${\cal L}_{VV}$ and the explicit chiral symmetry 
breaking term ${\cal L}_{\chi SB}$. In the MIT bag and the QMDD model, 
${\cal L}_M$ is replaced by the effective bag constant. 
At zero temperature, $\Omega$ can be expressed as
\begin{eqnarray}
\Omega &=& -\sum\limits_{i=u,d,s}\frac 1{8\pi^2}
\left\{\nu_i\left[\nu_i^2-m_i^{*2}\right]^{1/2}
\left[2\nu_i^2-5m_i^{*2}\right] \right .
\nonumber\\
&+& \left . 3m_i^{*4}{\rm ln}\left[\frac{\nu_i+\left(\nu_i^2-
m_i^{*2}\right)^{1/2}}{m_i^*}\right]\right\} - 
\frac{\mu_e^4}{12\pi^2}-{\cal L}_M,
\end{eqnarray}
where $\mu_e$ is the chemical potential of the electron and the quantity 
$\nu_i$ ($i$=$u$, $d$, $s$) is related to the usual chemical potential 
$\mu_i$ by $\nu_i=\mu_i-g_\omega^i\omega-g_\phi^i\phi$. The effective 
quark mass is given by $m_i^*=-g_\sigma^i\sigma-g_\zeta^i\zeta+m_{i0}$. 
The total baryon density is defined as 
\begin{equation}
\rho_B=\frac13 (\rho_u+\rho_d+\rho_s). 
\end{equation}
With the thermodynamical potential, the energy per volume 
$\varepsilon$ and pressure $p$ of the system can be derived as 
$\varepsilon=\Omega+\sum_{i=u, d, s, e} \mu_i \rho_i$ and $p=-\Omega$.

The mean field equations for the meson $\phi_i$ are obtained with 
$\frac{\partial\Omega}{\partial\phi_i}=0$. For the scalar mesons 
$\sigma$, $\zeta$ and $\chi$, the equations are expressed as 
\begin{eqnarray}\label{eq_gsigma} 
& &k_0\chi^2\sigma-4k_1\left(\sigma^2+\zeta^2\right)\sigma-2k_2\sigma^3 
-2k_3\chi\sigma\zeta-\frac{2\delta}{3\sigma}\chi^4 + 
\frac{\chi^2}{\chi_0^2} m_\pi^2f_\pi - \nonumber\\ 
&-&\left(\frac{\chi}{\chi_0}\right)^2m_\omega\omega^2 
\frac{\partial m_\omega}{\partial\sigma}
=\sum_{i=u,d}g_\sigma^i<\bar{\psi_i}\psi_i>, 
\end{eqnarray}
\begin{eqnarray}\label{eq_gzeta} 
& &k_0\chi^2\zeta-4k_1\left(\sigma^2+\zeta^2\right)\zeta-4k_2\zeta^3
-k_3\chi\sigma^2-\frac{\delta}{3\zeta}\chi^4 +\nonumber\\\
&+&\frac{\chi^2}{\chi_0^2}
\left(\sqrt{2}m_k^2f_k-\frac1{\sqrt{2}}m_\pi^2f_\pi\right) 
=\sum_{i=s}g_\zeta^i<\bar{\psi_i}\psi_i>,
\end{eqnarray}
\begin{eqnarray}\label{eq_chi} 
& &k_0\chi\left(\sigma^2+\zeta^2\right)-k_3\sigma^2
\zeta+\left(4k_4+1
+4{\rm ln}\frac{\chi}{\chi_0}-\frac{4\delta}{3}
{\rm ln}\frac{\sigma^2\zeta}{\sigma_0^2
\zeta_0}\right)\chi^3 + \nonumber \\
&+&\frac{2\chi}{\chi_0^2}\left[m_\pi^2f_\pi\sigma+ 
\left(\sqrt{2}m_k^2f_k-\frac1{\sqrt{2}}m_\pi^2f_\pi\right)\zeta\right] 
-\frac{\chi}{\chi_0^2}m_\omega^2\omega^2=0.
\end{eqnarray}
The equations for the vector mesons can be obtained in the same way as 
\begin{eqnarray}
\frac{\chi^2}{\chi_0^2}m_\omega^2\omega+4g_4\omega^3+12g_4\omega\rho^2
&=&\sum_{i=u,d}g_\omega^i<\bar\psi_i\gamma^0\psi_i>,\label{eq_gomega} \\
\frac{\chi^2}{\chi_0^2}m_\rho^2\rho+4g_4\rho^3+12g_4\omega^2\rho&=&
\sum_{i=u,d}g_\rho^i<\bar{\psi_i}\gamma_0\psi_i> , \label{eq_grho} \\
\frac{\chi^2}{\chi_0^2}m_\phi^2\phi+8g_4\phi^3&=&
\sum_{i=s}g_\phi^i<\bar{\psi_i}\gamma_0\psi_i> \label{eq_gphi} . 
\end{eqnarray}
The scalar and vector densities can be written as
\begin{eqnarray}\label{eq_rho} 
<\bar{\psi_i}\psi_i>&=&\frac6{(2\pi)^3}\int_0^{k_{F_i}} d^3k
\frac{m_i^*}{\sqrt{k^2+m_i^{*2}}} ,\\
<\bar{\psi_i}\gamma^0\psi_i>&=&\frac6{(2\pi)^3}\int_0^{k_{F_i}} d^3k 
\end{eqnarray}
with $k_{F_i}=\sqrt{\nu_i^2-m_i^{*2}}$. 

\section{Numerical results}

Now we investigate the implications of the previous formalism to 
strange quark matter. Compared to earlier applications to hadronic 
matter, here we need not to introduce the confining potential which 
combines quarks into baryons. In quark matter, the u, d and s quarks 
are deconfined. They only interact by scalar and vector mesons. The 
self-interactions between mesons are the same as in hadronic matter. 
All the parameters in this model are determined in our previous papers. 
They are listed in Table I. We assume that these parameters do not 
change when the model is applied to quark matter. In fact, the 
parameters which describe the interactions between fields 'should' be 
universal. The medium effects are included by the treatment 
(relativistic mean field theory) itself and are not included in the 
original Lagrangian. In our discussion, we first do not include the 
additional coupling ${\cal L}_h$ and discuss the effect of this 
term on strange quark matter later. 

If strange quark matter is stable and can survive for a long time, 
equilibrium with respect to the weak processes:
\begin{equation}
s\rightarrow u+e^-+\bar{\nu}_e, ~~~~ d\rightarrow u+e^-+\bar{\nu}_e, 
\end{equation}
may be achieved. 
If we neglect the chemical potential of the neutrino, 
the chemical potentials 
of quarks and electron have the following relation: 
\begin{equation}
\mu_d=\mu_s=\mu_u+\mu_e
\end{equation}
The values for $\mu_u$ and $\mu_d$ are determined by the baryon density 
and the total charge Q. As usually, we assume the charge Q to be zero. 
The equations for the mesons (\ref{eq_gsigma})-(\ref{eq_gphi}) can be 
solved simultaneously. 
The energy per baryon with different strength of vector coupling and for
different parameter sets are shown in Fig.1. When vector interactions 
are not considered, there is a local minimum of energy per baryon with 
parameter set A. The corresponding density is about $0.13fm^{-3}$. 
As shown later, at this density no strange quarks appear. When the 
vector interactions are included, even when the strength is half of that 
in hadronic matter, the local minimum for the energy per baryon disappears. 
The energy per baryon E/A increases monotonously with the increasing  
baryon density. 

In Fig.2, we show the fractions $r_i=\frac{\rho_i}{3\rho_B}$ of u, d, and s 
quark versus baryon density with and without vector interactions. 
The fraction of u quarks 
almost does not change with the density. There exists a relationship 
$\rho_d+\rho_s\simeq 2\rho_u$. Therefore, although we included the 
electron in our calculations, the fraction of electrons is very small. 
The chemical potential of the electron $\mu_e$ is not zero which results 
in a larger fraction of d than u quarks. At low 
density no strange quarks are present. The fraction of d quarks is 
about two times that of u quarks. Without vector interactions, when 
the density is larger than about $0.41fm^{-3}$, strange quarks appear. 
Compared to Fig.1, this density is larger than the density where the 
local minimum appears. This result is close to that of 
Ref. \cite{Buballa} where the SU(3) NJL model is used. Therefore, no 
stable strange quark matter can exist at zero pressure 
in contradiction to the original suggestion by Witten \cite{Witten}
The so-called strange star cannot be composed entirely by 
the deconfined u, d and s quarks. The stable strange quark matter can 
only exist in the core of these objects where the pressure is high 
enough to force the transition from hadronic matter to quark matter to 
occur. 

If strange quark matter is metastable, then it may be produced in heavy 
ion collisions. In this case, the $\beta$-equilibrium may not be 
achieved. In our calculations, we assume that $\mu_u=\mu_d=\mu_q$. 
The values of $\mu_q$ and $\mu_s$ are determined by the total baryon 
density and the strangeness fraction $f_s$ ($f_s=3r_s$). In Fig.3, we 
plot the effective masses of nonstrange and strange quarks versus the 
baryon density for different strangeness fraction with parameter set A. 
Both u (d) and s quark masses decrease with the increasing density. 
In nonstrange quark matter, the mass of the u (d) quark decreases more 
quickly than that of the s quark. When $f_s$ increases, the mass of the
s quark decreases at fixed density. At some high density, the strange 
quark mass is even lower than the mass of nonstrange quarks. Compared 
to the QMDD model, here the effective quark masses are obtained 
dynamically. The quark masses are not only density dependent but also 
strangeness fraction dependent which are not present in the QMDD model. 

In Figs.4-6 we plot the energy per baryon versus density for different 
values of $f_s$, which corresponds to different vector coupling constants. The 
solid and dashed lines are for parameter sets A and B, respectively. 
First we do not include the interactions between quarks and vector 
mesons. This is close to the QMDD model or SU(3) NJL model with only 
scalar type 4-quark interactions. 
In Fig.4, for parameter set A, there exists a local minimum of energy 
per baryon for any $f_s$. The baryon density at the minimum of energy 
per baryon first increases and then decreases when $f_s$ increases. 
For nonstrange quark matter, though there is a local minimum, the 
system has no positive binding energy compared to the vacuum mass of 
nonstrange quarks. For strange quark matter, the energy per baryon is 
lower than the masses of hyperons with the same strangeness number. When 
$f_s=1$, the binding energy is about 45 MeV compared to the vacuum 
constituent quark mass. The maximum binding energy is about 60 MeV and 
the corresponding $f_s$ is about 2.0. Metastable strange quark matter 
is therefore favored to have a large strangeness fraction (high 
negative charge). For the parameter set B, the system has a local 
minimum of energy per baryon only for some range of $f_s$, that is 
$1<f_s<3$. The maximum binding energy is about 5 MeV, when $f_s$ is 
around 2.0. 

When the interactions between quarks and vector mesons are included, 
the system is destabilized. In Fig.5, we plot the energy per baryon 
versus baryon density with the vector coupling constant half of that 
in hadronic matter. Compared to Fig.4, the energy per baryon becomes 
higher. For parameter set A, there exists a local minimum for 
$0<f_s<2$. The maximum binding energy is only about 10 MeV with 
$f_s \simeq$ 1.0. The corresponding density is also much lower 
compared to Fig.4. For parameter set B, there is no local minimum for 
any strangeness fraction. If the vector coupling is the same as in 
hadronic matter, as is shown in Fig.6, the energy per baryon will 
increase monotonously with the baryon density for both parameter 
sets A and B. The vector meson couplings are also discussed in 
Ref. \cite{Mishustin}. Though these two models are quite different, 
the results are comparable to each other. 

If vector interactions are included, even when the strength is only half 
of that in hadronic matter, the binding energy of metastable strange 
quark matter becomes small or negative. At the minimum of the energy 
per baryon, the pressure p of the system is zero. This kind of objects 
with finite volume are called strangelets. If the vector interactions 
are fully considered, the metastable strangelets can not be formed at 
zero pressure. 

Up to now, we did not consider the additional coupling between strange 
quark and scalar mesons ${\cal L}_h$, which is important to obtain 
reasonable hyperon potentials in hadronic matter \cite{Wang2}. When the additional 
term is included, the effective quark masses versus baryon density 
for parameter set A are given in Fig.7. The result for the strange quark 
mass evidently change, especially for large strangeness fraction, when 
compared to Fig.3. In nonstrange quark matter, the effective mass 
of s quarks almost stays constant. When the strangeness fraction 
is high, the strange quark has a lower mass when compared to Fig.3,
where ${\cal L}_h$ is not included

The effective quark masses will affect the energy of the system. In 
Fig.8 we plot the energy per baryon versus density with the vector 
coupling constant half of that in hadronic matter. The solid and dashed 
lines correspond to the cases without and with ${\cal L}_h$, 
respectively. For small strangeness fraction $f_s$, the results of 
these two cases are close. For large strangeness fraction, the 
additional term ${\cal L}_h$ will produce larger binding energy. If the 
vector coupling strength is as big as in hadronic matter, there 
is no local minimum for both cases, with or without ${\cal L}_h$. 
Therefore, the inclusion of the additional term ${\cal L}_h$ does not affect the 
main results of Fig.1. This is because when the $\beta$-equilibrium is 
achieved, the strangeness fraction of the system is smaller than 1.
The additional term ${\cal L}_h$ only gives sizable contributions for 
systems with a large strangeness fraction. 

\section{Conclusions}

We investigate strange quark matter in a chiral SU(3) quark mean field 
model. The effective quark masses are obtained dynamically by the quark 
meson interactions and they are both density and strangeness fraction 
dependent. The stability of strange quark matter is studied for 
different values of the vector coupling constant and for different 
parameter sets. The effect of the additional term ${\cal L}_h$ is also 
discussed.

If the strange quark is stable and can survive a long time, the 
$\beta$-equilibrium can be achieved. The strangeness fraction $f_s$ is 
smaller than 1. In the chiral SU(3) quark model, at the density where 
the system has a local minimum for the energy per baryon, no strange 
quarks appear. Even without the vector meson coupling, this nonstrange 
quark matter has a negative binding energy and cannot be bound. 
Therefore, opposed to the suggestion by Witten, stable quark matter at 
zero pressure cannot exist even without vector meson interactions. 

If quark matter can be produced in heavy ion collisions, the 
$\beta$-equilibrium may not be achieved. This metastable strange quark 
matter can have a high strangeness fraction (high negative charge). 
When we do not take the vector meson interaction into account, as in 
the QMDD model or SU(3) NJL model with only scalar type 4-quark 
interaction, the maximum binding energy is about 5 MeV - 60 MeV and 
the corresponding strangeness fraction is about 1 - 2. If we assume 
that the strength of vector meson coupling is half of that in hadronic 
matter (in this case, the strength of vector coupling is comparable 
to the one of Ref. \cite{Mishustin}), the maximum binding energy 
decreases. Inclusion of the additional term ${\cal L}_h$ 
will forces the system to have a larger strangeness fraction and 
binding energy. When the vector coupling constant is the same as for
hadronic matter, no metastable strange quark matter (strangelets) can 
be formed in heavy ion collisions. 

However, strange quark matter can still exist in the core of a neutron 
star. The phase transition from hadronic matter to quark matter can 
occur at high density and pressure which is caused by gravity. 
Because both hadronic and quark matter can be described in 
the SU(3) quark mean field model, it is of interest to investigate this 
phase transition in this model. This will be studied in the future.

\vspace*{.3cm}

{\bf Acknowledgements}. P. Wang would like to thank the Institute for 
Theoretical Physics, University of T\"ubingen for their hospitality. 
This work was supported by the Alexander von Humboldt Foundation and 
by the Deutsche Forschungsgemeinschaft (DFG) 
under contracts FA67/25-1, GRK683.

\newpage 

\begin{table}[p]
\caption{Parameters of sets A and B of the model.}
\begin{center}
\begin{tabular}{|c|c|c|c|c|c|c|c|c|c|c|}
set & $k_0$ &$k_1$ &$k_2$ &$k_3$ & $k_4$ &$g_s$&$g_v$ 
&$g_4$ &$h_1$ & $h_2$\\ \hline
A & 4.94 & 2.12 & -10.16 & -5.38 & -0.06 & 4.76 & 10.92 & 37.5 & -2.20 
& 3.24 \\
B & 3.83 & 2.64 & -10.16 & -3.40 & -0.18 & 4.76 & 10.13 & 0.0  & -2.03 
& 2.55 \\
\end{tabular}
\end{center}
\end{table}
 
\centerline{Figure captions}

{\bf Fig.1:} Energy per baryon versus baryon density for different 
vector coupling constants. The solid and dashed lines are for parameter 
sets A and B. Results are shown for the case of $\beta$-equilibrium.

\bigskip 

{\bf Fig.2:} Fractions of u, d and s quarks in strange matter versus 
baryon density with and without vector interactions, respectively.
The solid and dashed lines are for parameter sets A and B. 
Results are shown for the case of $\beta$-equilibrium.
\bigskip 

{\bf Fig.3:} The effective nonstrange and strange quark mass versus baryon 
density for different strangeness fraction. Parameters are taken from set A.

\bigskip 

{\bf Fig.4:} The energy per baryon versus baryon density for different 
strangeness fraction for parameter sets A and B. The interaction 
between quarks and vector mesons is not considered. 

\bigskip 

{\bf Fig.5:} Same as in Fig. 4, but here the quark-vector meson interaction
is included. The value of the vector coupling constant is half of that used
in hadronic matter.

\bigskip 

{\bf Fig.6:} Same as Fig. 5, but the vector coupling constant is of the strength
as in hadronic matter. 

\bigskip 

{\bf Fig.7:} Same as Fig. 3, but the the additional term ${\cal L}_h$ is 
included in the calculation. 

\bigskip 

{\bf Fig.8:} The energy per baryon versus baryon density calculated with 
parameter set A. The vector coupling constant is half of that in 
hadronic matter. Solid and dashed lines correspond to the case 
without and with the additional term ${\cal L}_h$, respectively. 

\newpage 

\begin{figure}
\centering{\
\epsfig{figure=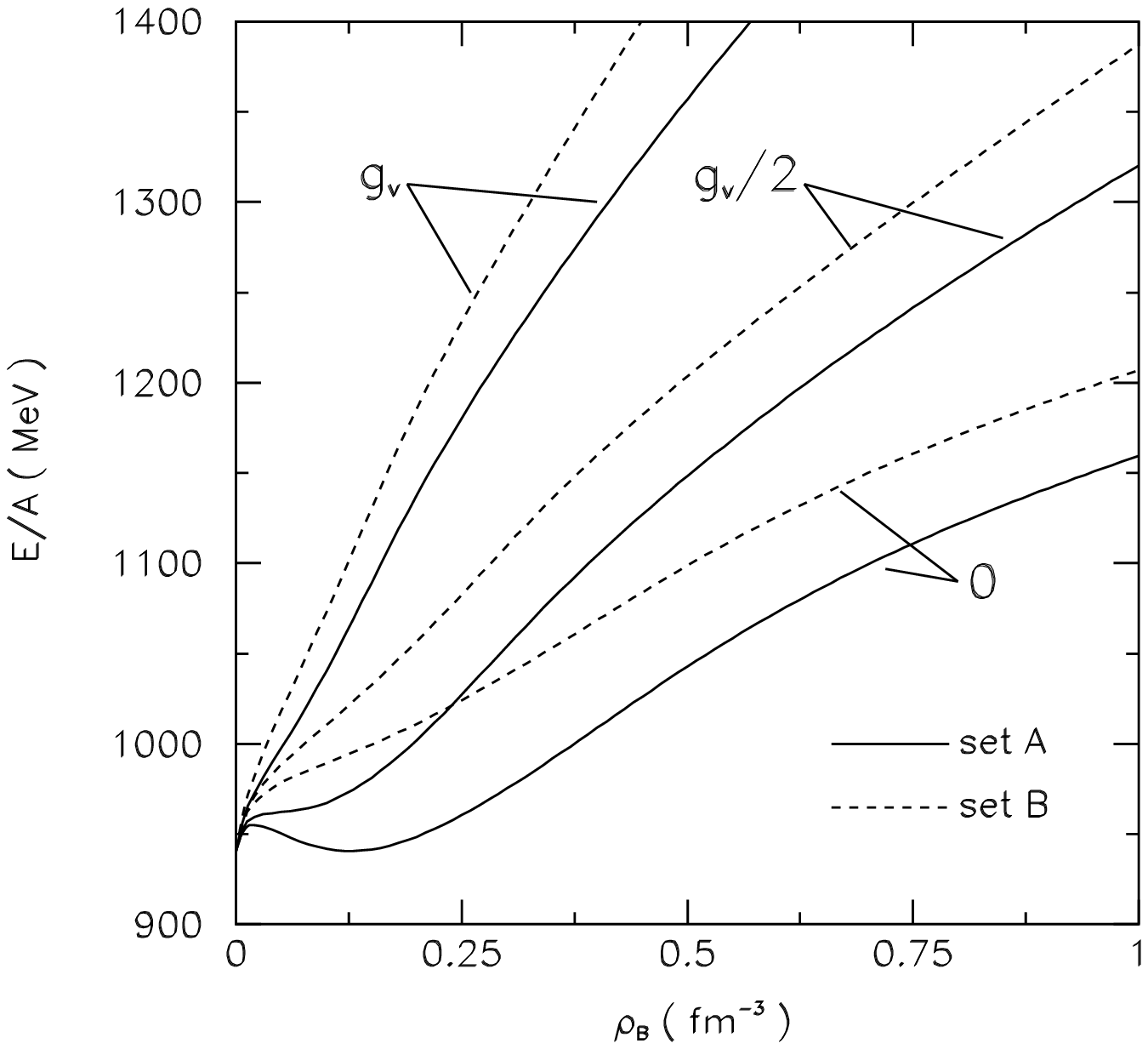,height=21cm}
}
\end{figure}
\vspace*{-6cm}
\centerline{\bf Fig.1}

\newpage 

\begin{figure}
\centering{\
\epsfig{figure=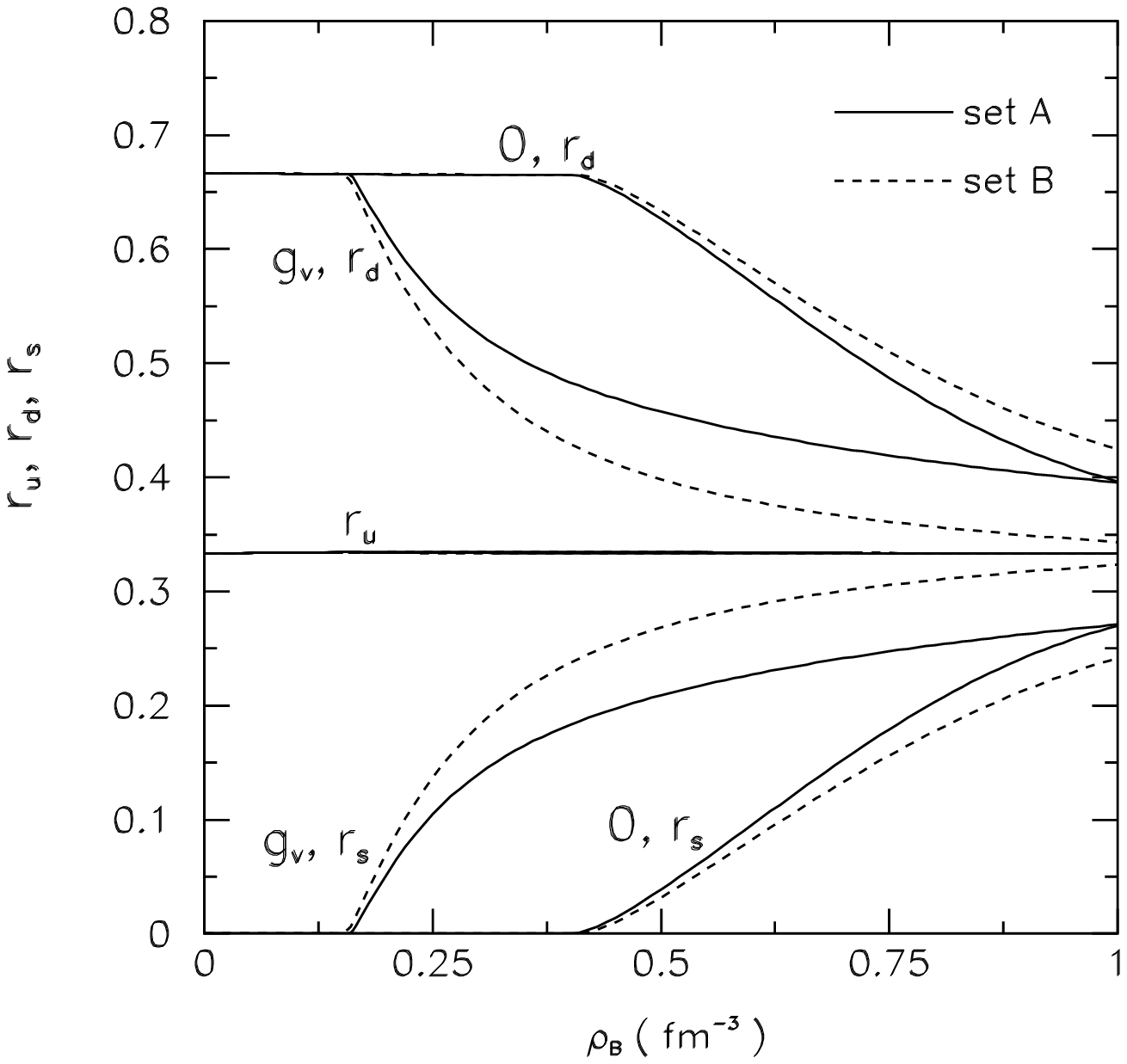,height=21cm}
}
\end{figure}
\vspace*{-6cm}
\centerline{\bf Fig.2}

\newpage 

\begin{figure}
\centering{\
\epsfig{figure=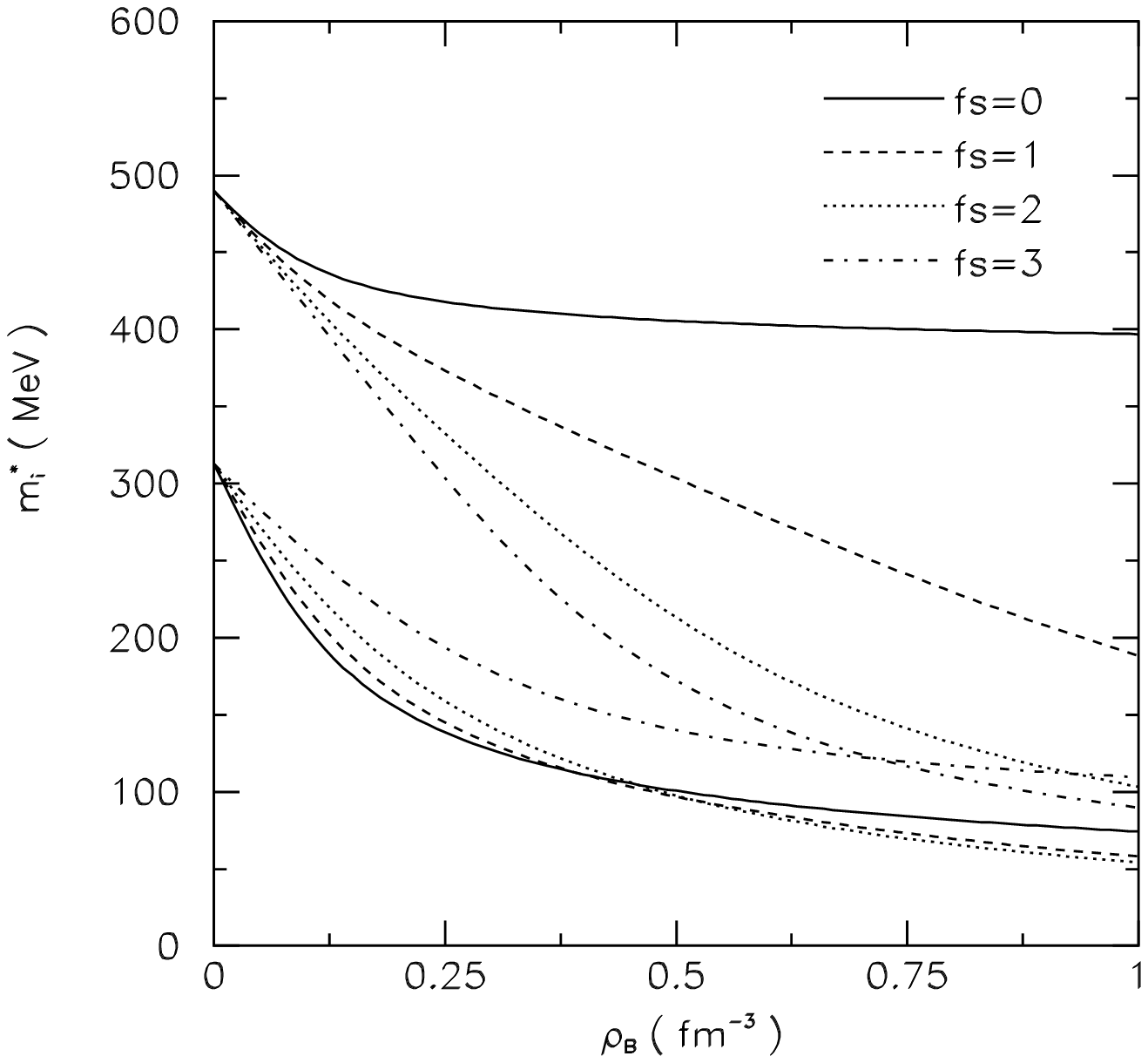,height=21cm}
}
\end{figure}
\vspace*{-6cm}
\centerline{\bf Fig.3}

\newpage 

\begin{figure}
\centering{\
\epsfig{figure=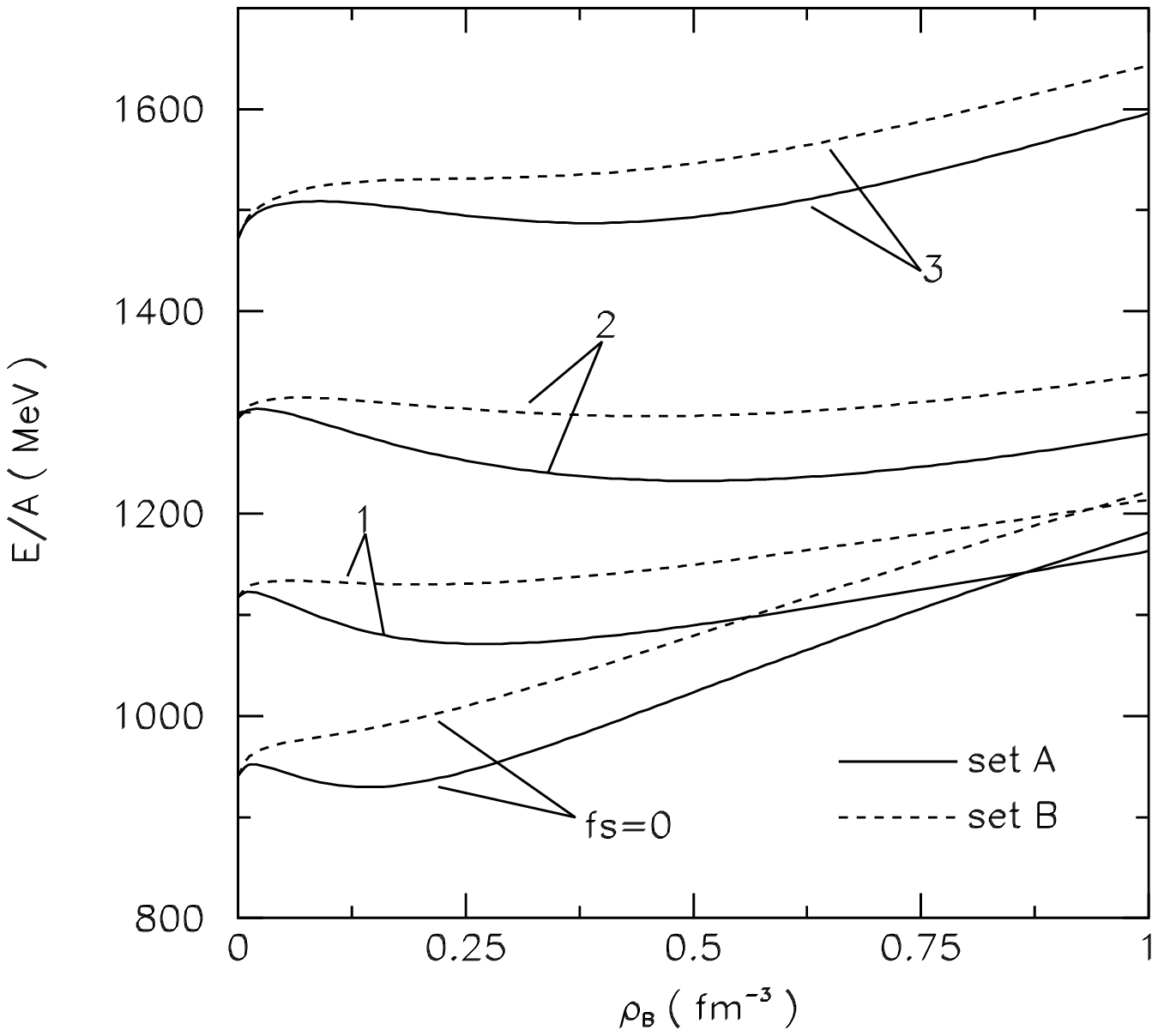,height=21cm}
}
\end{figure}
\vspace*{-6cm}
\centerline{\bf Fig.4}

\newpage 

\begin{figure}
\centering{\
\epsfig{figure=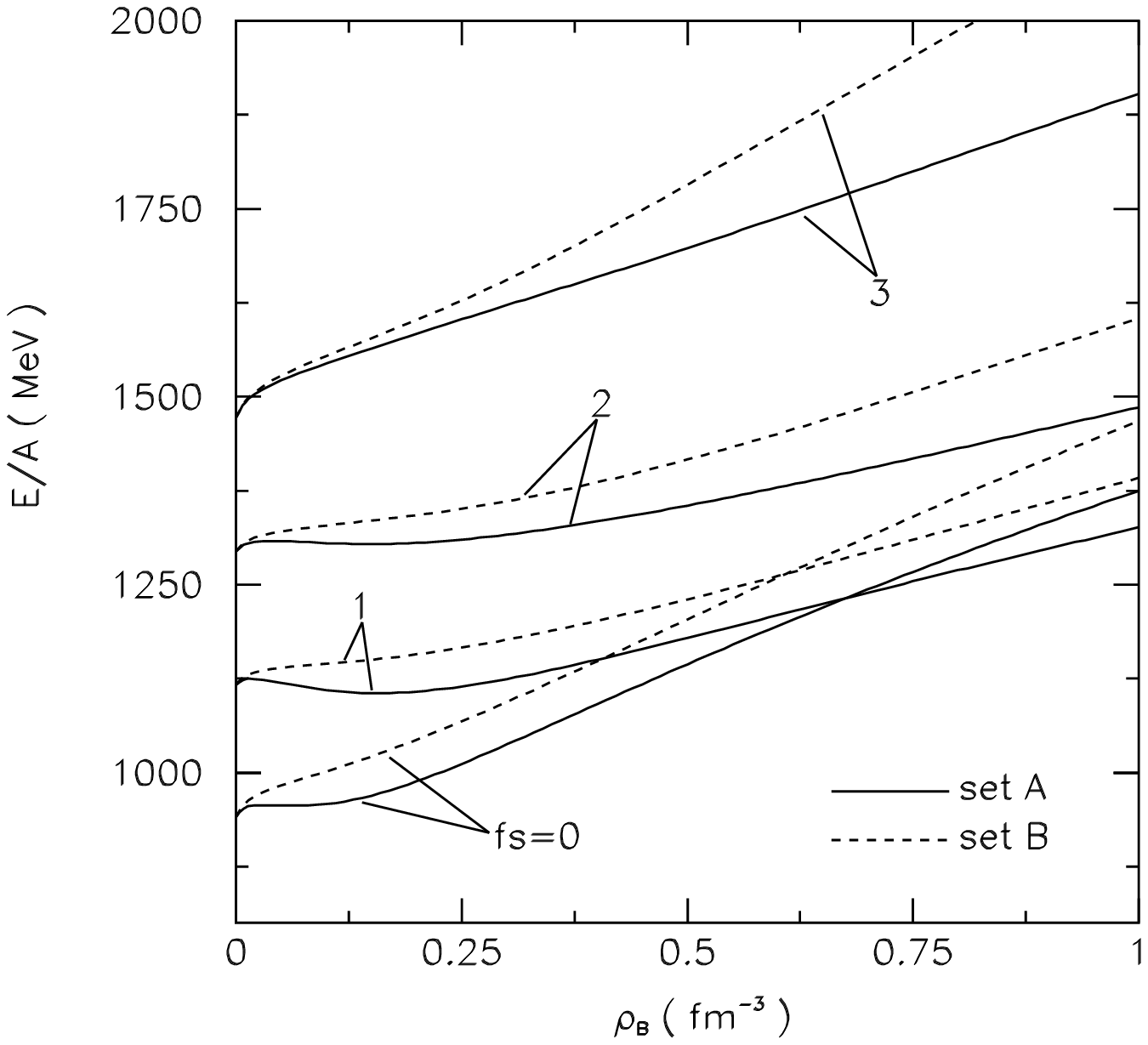,height=21cm}
}
\end{figure}
\vspace*{-6cm}
\centerline{\bf Fig.5}

\newpage 

\begin{figure}
\centering{\
\epsfig{figure=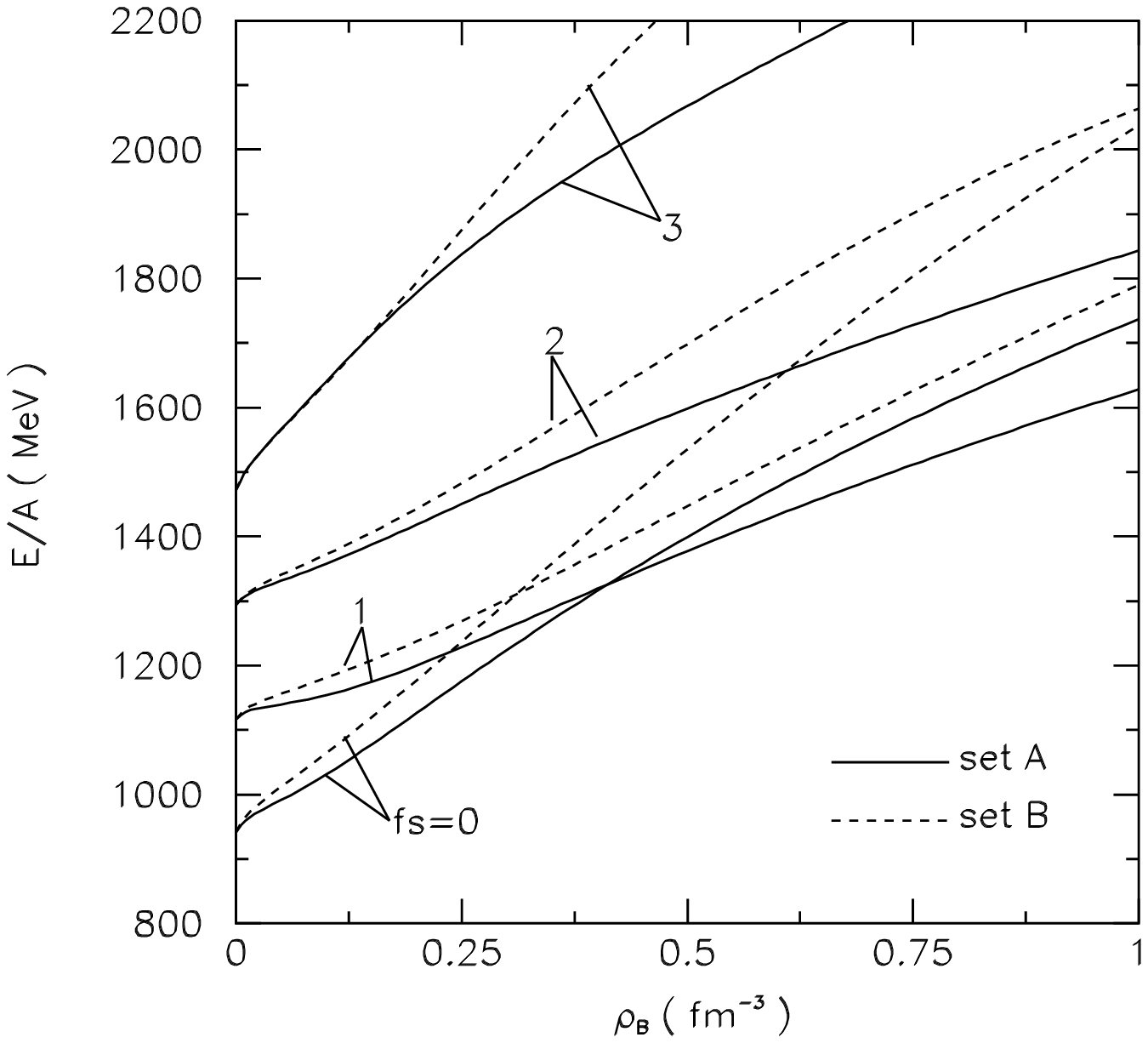,height=21cm}
}
\end{figure}
\vspace*{-6cm}
\centerline{\bf Fig.6}

\newpage 

\begin{figure}
\centering{\
\epsfig{figure=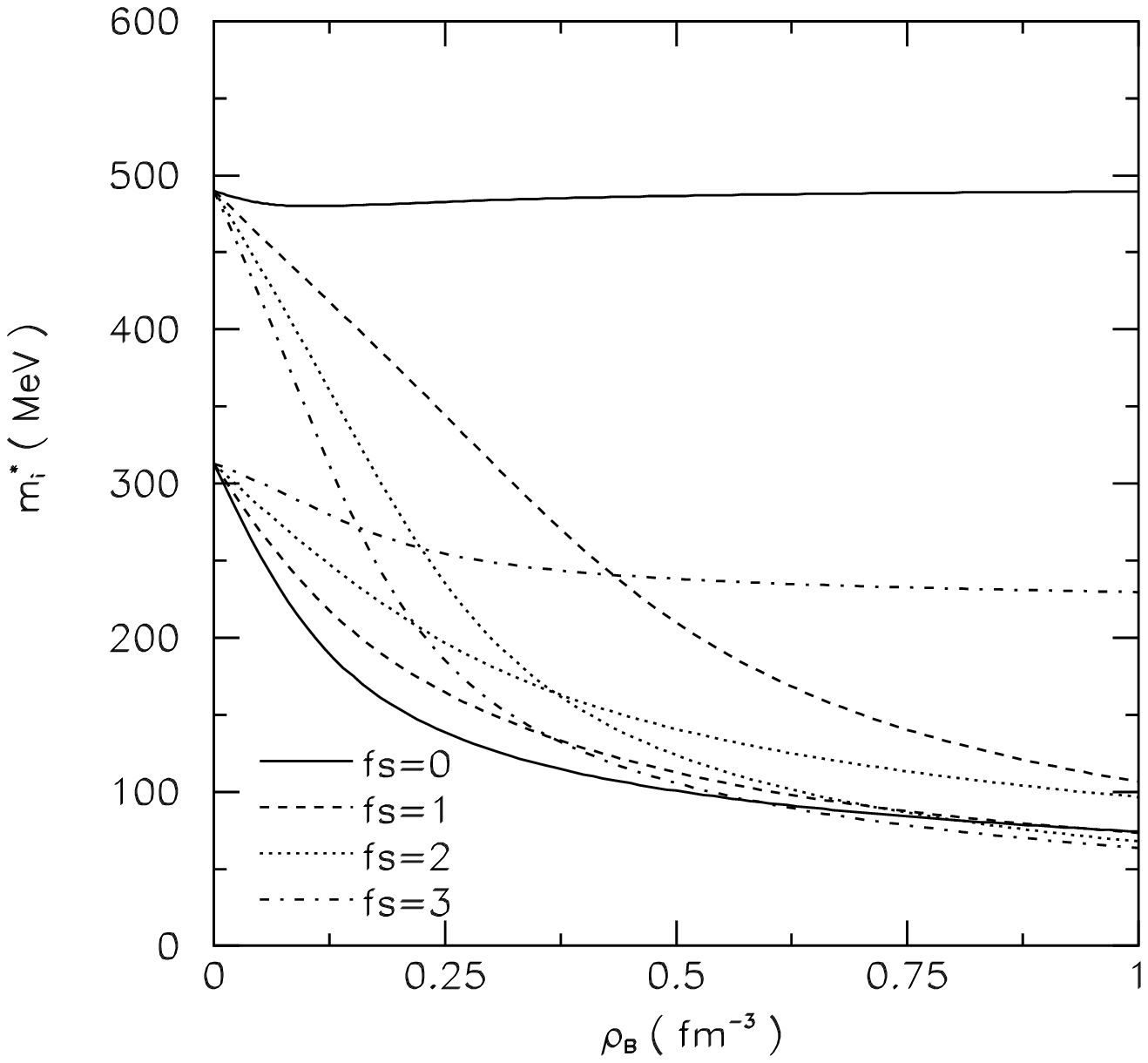,height=21cm}
}
\end{figure}
\vspace*{-6cm}
\centerline{\bf Fig.7}

\newpage 

\begin{figure}
\centering{\
\epsfig{figure=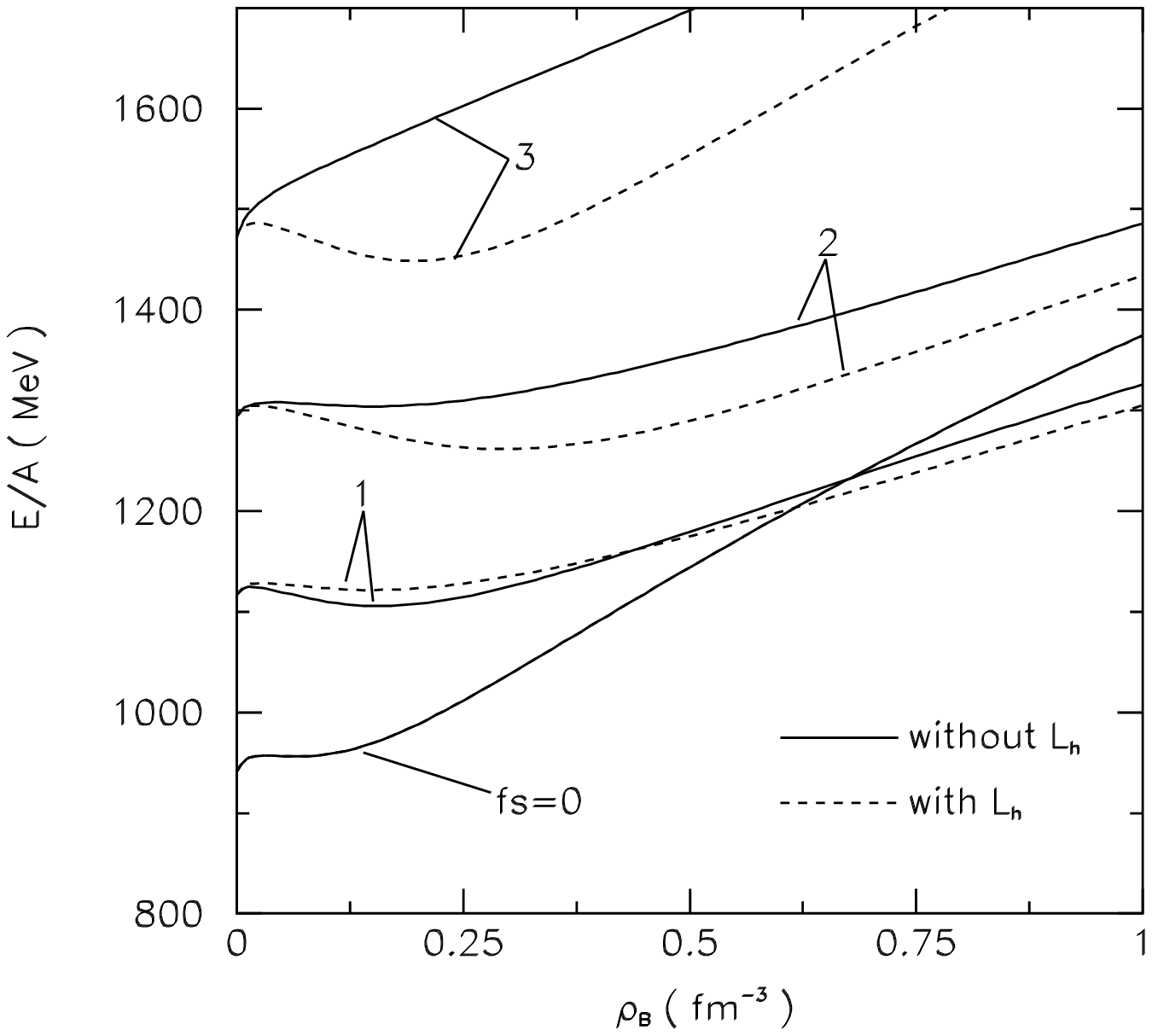,height=21cm}
}
\end{figure}
\vspace*{-6cm}
\centerline{\bf Fig.8}


\begin{thebibliography}{999} 
\bibitem{Witten}E.~Witten, Phys. Rev. D {\bf 30}, 272 (1984). 
\bibitem{Kasuya}M.~Kasuya {\it et al.}, Phys. Rev. D {\bf 47}, 
2153 (1993). 
\bibitem{Ichimura}M.~Ichimura {\it et al.}, Nuovo Cim. A {\bf 106}, 
843 (1993). 
\bibitem{Capdevielle}J.~N.~Capdevielle {\it et al.}, Nuovo Cim. C 
{\bf 19}, 623 (1996). 
\bibitem{Bombaci}I.~Bombaci, Phys. Rev. C {\bf 55}, 1587 (1997). 
\bibitem{Cheng}K.~S.~Cheng, Z.~G.~Dai, D.~M.~Wai, and T.~Lu, 
Science {\bf 280}, 407 (1998). 
\bibitem{Dey}M.~Dey, I.~Bombaci, J.~Dey, S.~Ray, and B.~C.~Samanta, 
Phys. Lett. B {\bf 438}, 123 (1998); Erratum Phys. Lett. B {\bf 467}, 
303 (1999). 
\bibitem{Li}X.~D.~Li, I.~Bombaci, M.~Dey, J.~Dey, and E.~P.~J.~van den 
Heuvel, Phys. Rev. Lett. {\bf 83}, 3776 (1999). 
\bibitem{Greiner1}C.~Greiner, P.~Koch, and H.~St\"ocker, Phys. Rev. 
Lett. {\bf 58}, 1825 (1987). 
\bibitem{Greiner2}C.~Greiner, D.~H.~Rischke, H.~St\"ocker, and P.~Koch, 
Phys. Rev. D {\bf 38}, 2797 (1988). 
\bibitem{Greiner3}C.~Greiner, J.~Schaffner, and H.~St\"ocker, 
Nucl. Phys. Proc. Suppl. B {\bf 24}, 239 (1991). 
\bibitem{Barrette}J.~Barrette {\it et al.}, Phys. Lett. B {\bf 252}, 
550 (1990); M.~Aoki {\it et al.}, Phys. Rev. Lett. {\bf 69}, 
2345 (1992); K.~Borer {\it et al.}, {\it ibid}. {\bf 72}, 1415 (1994); 
D.~Beavis {\it et al.}, {\it ibid} {\bf 75}, 3078 (1995). 
\bibitem{Ardouin}D.~Ardouin {\it et al.}, Phys. Lett. B {\bf 446}, 
191 (1999). 
\bibitem{Armstrong}T.~A.~Armstrong {\it et al.}, Phys. Rev. C {\bf 59}, 
1829 (1999); Phys. Rev. C {\bf 63} 054903 (2001). 
\bibitem{Chodos}A.~Chodos {\it et al.}, Phys. Rev. D {\bf 9}, 
3472 (1974). 
\bibitem{Farhi}E.~Farhi and R.~L.~Jaffe, Phys. Rev. D {\bf 30}, 
2379 (1984). 
\bibitem{Berger}M.~S.~Berger and R.~L.~Jaffe, Phys. Rev. C {\bf 35},  
213 (1987). 
\bibitem{Madsen}J.~Madsen, Phys. Rev. Lett. {\bf 70}, 391 (1993); 
Phys. Rev. D {\bf 47}, 5156 (1993). 
\bibitem{Ukawa}A.~Ukawa, Nucl. Phys. A {\bf 498}, 227c (1989). 
\bibitem{Fowler}G.~N.~Fowler, S.~Raha, and R.~M.~Weiner, 
Z. Phys. C {\bf 9}, 271 (1981). 
\bibitem{Chakrabarty}S.~Chakrabarty, S.~Raha, and B.~Sinha, 
Phys. Lett. B {\bf 229}, 112 (1989).
\bibitem{Chakrabarty2}S.~Chakrabarty, Phys. Rev. D {\bf 43}, 
627 (1991). 
\bibitem{Chakrabarty3}S.~Chakrabarty, Phys. Rev. D {\bf 48}, 
1409 (1993). 
\bibitem{Benvenuto}O.~G.~Benvenuto and G.~Lugones, Phys. Rev. 
D {\bf 51}, 1989 (1995). 
\bibitem{Peng}G.~X.~Peng, H.~C.~Chiang, P.~Z.~Ning, and B.~S.~Zou, 
Phys. Rev. C {\bf 59}, 3452 (1999). 
\bibitem{Wang1}P.~Wang, Phys. Rev. C {\bf 62}, 015204 (2000). 
\bibitem{Alberico}W.~M.~Alberico, A.~Drago, and C.~Ratti, 
hep-ph/0110091. 
\bibitem{Jaminon}M.~Jaminon and B.~Van den Bossche, Nucl. Phys. 
A {\bf 686}, 341 (2001). 
\bibitem{Mishustin}I.~N.~Mishustin, L.~M.~Satarov, H.~St\"ocker, and 
W.~Greiner, Phys. Atom. Nucl. {\bf 64}, 802 (2001). 
\bibitem{Buballa}M.~Buballa and M.~Oertel, Phys. Lett. 
B {\bf 457}, 261 (1999). 
\bibitem{Wang2}P.~Wang, Z.~Y.~Zhang, Y.~W.~Yu, R.~K.~Su, and 
H.~Q.~Song, Nucl. Phys. A {\bf 688}, 791 (2001). 
\bibitem{Wang3}P.~Wang, H.~Guo, Z.~Y.~Zhang, Y.~W.~Yu, R.~K.~Su,   
and H. Q. Song, Nucl. Phys. A {\bf 705}, 455 (2002). 
\bibitem{Papazoglou}P. Papazoglou, D. Zschiesche, S. Schramm, 
J.Schaffner-Bielich,  H. St\"ocker, and W. Greiner, Phys. Rev. 
C {\bf 59}, 411 (1999). 
\end{thebibliography}
\end{document}